\documentclass{appolb}
\pdfoutput=1
\usepackage{epsfig}
\usepackage{delarray,amsmath,bbm,slashed}

\newcommand{\nn}{\nonumber}
\newcommand{\beq} {\begin{equation}}
\newcommand{\eeq} {\end{equation}}
\newcommand{\beqa} {\begin{eqnarray}}
\newcommand{\eeqa} {\end{eqnarray}}
\newcommand{\mrm}[1] {{\mathrm{#1}}}

\newcommand{\gev}{{\mrm{GeV}}}

\newcommand{\lqcd}{\Lambda_{QCD}}

\newcommand{\ieps}{i\varepsilon}

\newcommand{\order}[1]{${\cal O}\left(#1 \right)$}
\newcommand{\morder}[1]{{\cal O}\left(#1 \right)}
\newcommand{\eq}[1]{(\ref{#1})}
\newcommand{\fig}[1]{Fig.~\ref{#1}}

\newcommand{\inv}[1]{\frac{1}{#1}}
\newcommand{\ket}[1]{\vert{#1}\rangle}
\newcommand{\bra}[1]{\langle{#1}\vert}

\newcommand{\bs}[1]{\boldsymbol{#1}}

\newcommand{\qu}{{\rm q}}
\newcommand{\qb}{{\rm\bar q}}

\newcommand{\kt}{\bs{k}_\perp}
\newcommand{\pt}{\bs{p}_\perp}

\newcommand{\halft}{{\textstyle \frac{1}{2}}}
\newcommand{\gsim}{\buildrel > \over {_\sim}}

\begin{document}
\title{QCD factorization at fixed Q$^2$(1-x)\ %
\thanks{Talk at Epiphany meeting in commemoration of Jan Kwieci\'nski, Krakow,
 January 2009. Based on work done in collaboration with Matti J\"arvinen and Samu Kurki \cite{Hoyer:2008fp}.}%
}
\author{Paul Hoyer
\address{Department of Physics and Helsinki Institute of Physics\\
              POB 64, FIN-00014 University of Helsinki, Finland}
}
\maketitle
\begin{abstract}
Amplitudes of hard {\it exclusive} processes such as $\gamma^*(Q^2) N \to \gamma Y$, where $Y=N$ (DVCS) or any other state with a limited mass ($M_Y^2 \ll Q^2$), factorize into a hard subprocess amplitude and a target (transition) GPD. The corresponding {\it inclusive} cross section, summed over all states $Y$ of a given (limited) mass, is then given by the discontinuity of a forward multiparton distribution. An application to the Drell-Yan process $\pi^+ N \to \gamma^*(x_F,Q^2)+Y$ allows to explain the observed longitudinal polarization of the virtual photon at high $x_{F}$.
\end{abstract}

\vspace{-10cm}
\hfill {HIP-2009-07/TH}
\vspace{10cm}
  
\section{The inclusive -- exclusive connection}

\begin{figure}[h]
\centerline{\epsfig{file=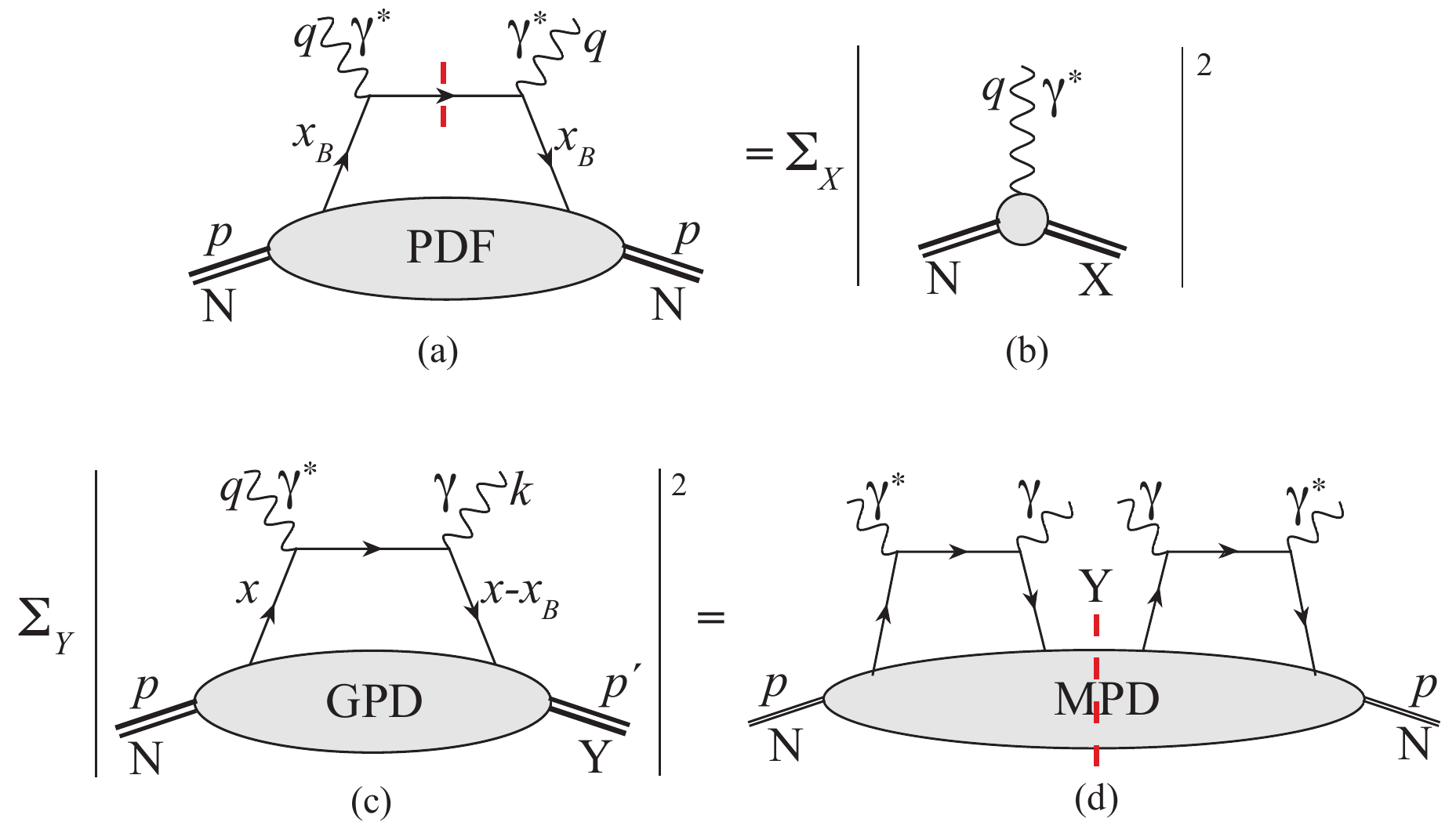,width=0.9\columnwidth}}
\caption{The discontinuity of the forward $\gamma^* N \to \gamma^* N$ amplitude (a) equals the DIS cross section (b). The PDF is real, hence the discontinuity is obtained by cutting only the struck quark (dashed line). Analogously, the sum over $Y$ of the squares of the $\gamma^* N \to \gamma Y$ amplitudes (c) is given by the discontinuity of the forward multiparton distribution (d).
 \label{DIS}}
\end{figure}

The optical theorem expresses the cross section of deep inelastic lepton scattering $eN\to eX$ as a discontinuity of the forward $\gamma^*(q)N(p)$ amplitude. In the Bjorken limit where the photon virtuality $-q^2=Q^2\to\infty$ at fixed $x_B=Q^2/2p\cdot q$ this amplitude factorizes into the $\gamma^*(q)\qu(x_B p) \to \gamma^*(q)\qu(x_B p)$ hard subprocess amplitude and a target parton distribution (PDF, \fig{DIS}(a)). 

The factorization property can be readily understood intuitively. The invariant mass $M_X$ of the inclusive system grows with $Q$,
\beq\label{cmen}
M_X^2 = (p+q)^2 = m_N^2+ \inv{x_B}(1-x_B)Q^2 \to \infty \hspace{.5cm} {\rm in\ the\ Bj\ limit}
\eeq
The virtual photon transfers its large momentum to the struck quark which hadronizes nearly independently of the target spectators. Factorization does not hold at low hadronic mass $M_X$ (\ie, for $x_B \to 1$) due to coherence effects between the struck quark and the spectators. The optical theorem itself is exact and may be applied even for $X=N$ ($x_B=1$), in which case the discontinuity of the forward $\gamma^* N \to \gamma^* N$ amplitude measures the square of the elastic form factor in \fig{DIS}(b). Intriguing similarities -- Bloom-Gilman duality \cite{Melnitchouk:2005zr} -- are nevertheless observed between the nucleon elastic and transition form factors on the one hand and the factorized, high $M_X^2$ DIS cross section on the other. This may contain clues to the dynamics of the exclusive form factors \cite{Hoyer:2007ns}.

Instead of the discontinuity of the forward $\gamma^* N \to \gamma^* N$ amplitude we may consider the non-forward $\gamma^* N \to \gamma Y$ amplitude measured in Deeply Virtual Compton Scattering (DVCS). As indicated in Fig.~\ref{DIS}(c) this amplitude factorizes similarly to DIS in the Bjorken limit. The large momentum imparted to the struck quark is carried away by the final photon and the quark then fuses with the target spectators into a low-mass system $Y$, often taken to be the nucleon itself ($Y=N$). The soft target dynamics is described by a Generalized Parton Distribution (GPD) \cite{Diehl:2003ny} which is real and in the forward limit reduces to the PDF of Fig.~\ref{DIS}(a). An integral over $x$ of the GPD gives the elastic (or transition, $N \to Y$) nucleon form factors. These form factors could not be obtained from the PDF of Fig.~\ref{DIS}(a) since DIS factorization breaks down at fixed $M_X$, \ie, for $x_B\to 1$ with $(1-x_B)Q^2$ fixed.

The GPD factorization shown in Fig.~\ref{DIS}(c) works in the Bj limit for any final state $Y$ whose mass is small compared to the total CM energy given in \eq{cmen}, $M_Y^2 \ll M_X^2$. $M_Y$ is kinematically constrained by the momentum $k$ carried away by the real photon. We may parametrize the external momenta using $p=(p^+,p^-,\pt)$ notation ($p^\pm=p^0\pm p^3$) as
\beqa
q&=&(-Q,Q,\bs{0}_\perp)\nn\\
p&=&(Q/x_B,m_N^2 x_B/Q,\bs{0}_\perp)\\
k&=&({k}_\perp^{2}/(x_F Q),x_FQ,\kt)\nn
\eeqa
where $k_\perp \ll Q$ is the  transverse momentum of the final photon and the Feynman $x_F = k^-/q^-$. This gives
\beq
M_Y^2 = (p+q-k)^2 = \frac{1-x_B}{x_B}(1-x_F)Q^2\,\left[1+\morder{\frac{1}{Q^2}}\right]
\eeq
GPD factorization works at any fixed $M_Y$, \ie, keeping $(1-x_F)Q^2$ fixed. We may then use completeness in the system $Y$ to relate the {\it inclusive} DVCS process $\gamma^* N \to \gamma Y$ to the discontinuity of the forward multiparton distribution (MPD) shown in \fig{DIS}(d).

\section{The BB limit}

The method illustrated above for DVCS may be applied to many other processes. We were motivated \cite{Hoyer:2008fp} particularly by the data on the Drell-Yan reaction $\pi^+N \to \gamma^*Y$, which may be viewed as a time reversed version of DVCS, with the real photon replaced by the pion. A dramatic change in the polarization of the virtual photon, from transverse to longitudinal, was observed \cite{Anderson:1979xx} at high $x_F$. According to an early analysis by Berger and Brodsky \cite{Berger:1979du} this signals the emergence of a dynamics in which both valence quarks of the pion scatter coherently, transferring nearly all their momentum $(x_F\to 1)$ and helicity $(\lambda=0)$ to the virtual photon. Thus we refer to the limit considered here as the
\beq\label{bblimit}
{\rm BB\ limit:}\ \ \ Q^2 \to \infty \ \ \ {\rm at\ fixed}\ \ Q^2(1-x)
\eeq
Here $x$ may refer either to the momentum fraction $x_F$ of particle in the final state (such as the real photon in DVCS) or to a parton momentum fraction in a hadron (such as a valence quark in the pion of the Drell-Yan process) and $Q$ is the hard scale (a large virtuality or transverse momentum).

The life-time of a hadron Fock state is inversely proportional to $\Delta E$, the energy difference between the hadron and its Fock state. At high hadron momentum $p$, 
\beq\label{endiff}
2p\Delta E \simeq m_h^2-\sum_i\frac{p_{i\perp}^2+m_i^2}{x_i}
\eeq
where the $x_i$ are the momentum fractions and $p_{i\perp}$ the transverse momenta of the partons in the Fock state. In the BB limit \eq{bblimit} a parton with $x \to 1$ and $p_\perp^2 \sim Q^2$ thus contributes to $\Delta E$ similarly as the partons carrying $1-x \sim \lqcd^2/Q^2$ and $p_\perp^2 \sim \lqcd^2$. Hence soft interactions of the partons with low $x$ are coherent with, and influence, the hard interactions of the large $x$ parton.

\begin{figure}[h]
\centerline{\epsfig{file=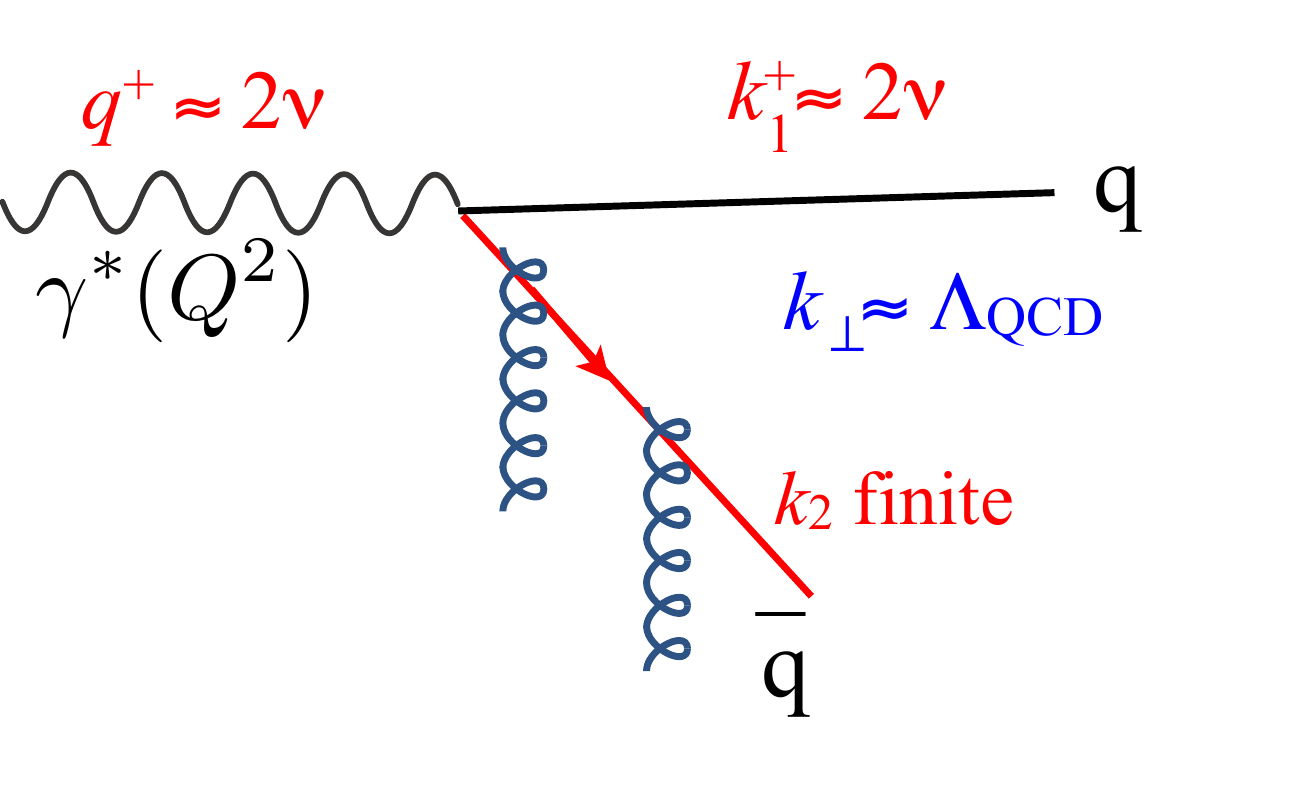,width=0.5\columnwidth}}
\caption{Light-Front time $x^+$ development of DIS when the virtual photon momentum is along the positive $z$-axis. The photon fluctuates into an asymmetric $\qu\qb$ pair where the quark carries nearly all the momentum whereas the antiquark has finite momentum in the target rest frame even as $q^+ \to \infty$. Soft scattering of the antiquark in the target (indicated by the vertical gluons) triggers the hard DIS process.
\label{dipole}}
\end{figure}

A good example of such coherence is provided by DIS itself, viewed in the ``target rest frame'' ($q^+ \simeq 2\nu$). The Light-Front (LF) time ($x^+$) development is sketched in \fig{dipole}. The virtual photon splits asymmetrically into a quark pair, $\gamma^* \to \qu(z)+\qb(1-z)$, with the quark carrying nearly all the momentum ($k_{\qu}^+ \simeq 2\nu$) while the antiquark momentum $k_{\qb}^+ = 2\nu(1-z) \sim \lqcd$ is fixed as $\nu\to\infty$. Since $\nu \propto Q^2$ the $\qu\qb$ Fock state of the virtual photon illustrates the BB limit \eq{bblimit}. In light-cone gauge ($A^-=0$) only the $\qb$ scatters (softly) in the target, which sets the quark on-shell and thus ``causes'' the hard DIS interaction. The hard $\gamma^*$ and soft $\qb$ interactions are coherent due to their commensurate lifetimes: $x_{\qb}^+ \sim 1/k_{\qb}^-$ and $x_{\gamma^*}^+ \sim 2\nu/Q^2=1/mx_B$ are both finite. In the usual ``handbag'' picture of DIS the antiquark in \fig{dipole} is viewed as the target quark which is struck by the $\gamma^*$, and its soft target interactions are part of the bound state dynamics.

\begin{figure}[h]
\centerline{\epsfig{file=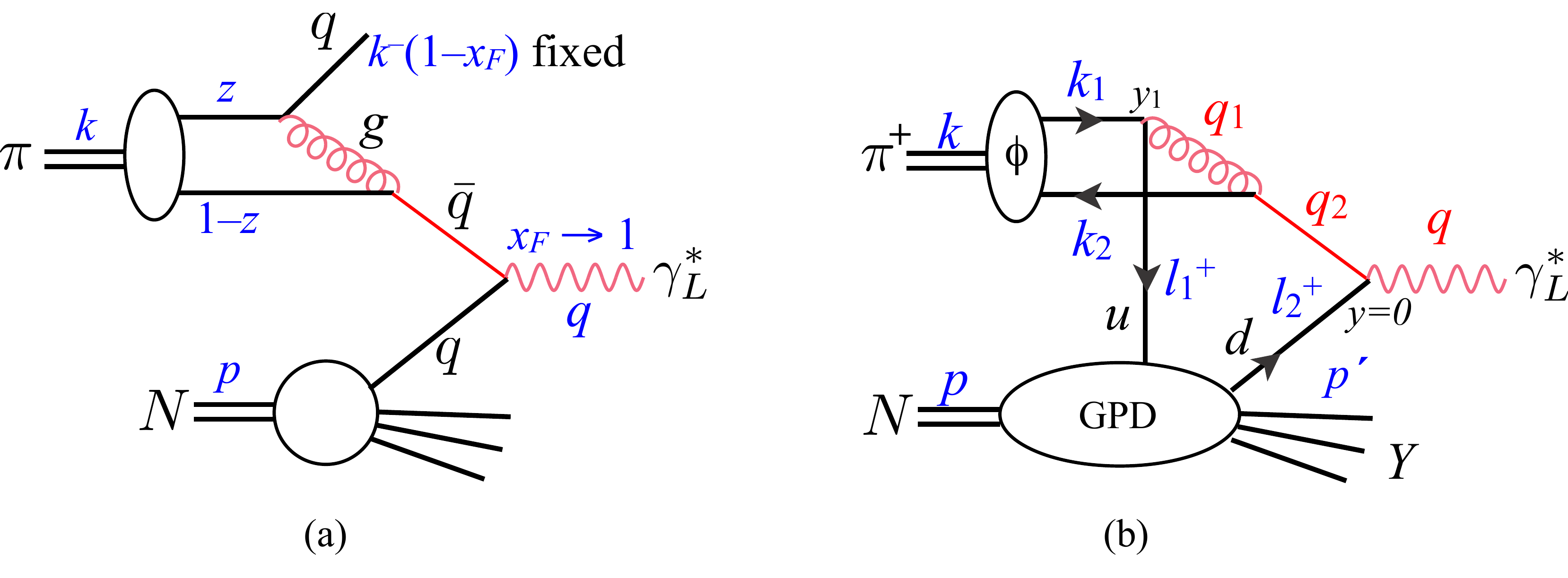,width=1.\columnwidth}}
\caption{(a) For the photon to carry nearly all the momentum of the pion ($x_F\to 1$) the momentum $zk$ carried by the quark $\qu$ is transferred to the antiquark $\qb$ via gluon $g$ exchange, with both the $g$ and the $\qb$ acquiring virtualities of \order{Q^2}. (b) The gluon emission leaves the quark with a finite momentum $\ell_1^-=k^-(1-x_{F})$ in the target rest frame. The interactions of the quark within the target are coherent with the hard subprocess and described by a target (transition) GPD, which involves an integral over $\ell_1^-$ and $\ell_{1\perp}$.
\label{dybb}}
\end{figure}

\section{Drell-Yan in the BB limit}

The dynamics of $\pi^+ N \to \gamma^*(x_F)+Y$ for
\beq\label{xf}
x_F = \frac{q^-}{k^-} \to 1
\eeq
as discussed by Berger and Brodsky \cite{Berger:1979du} is shown in \fig{dybb}(a). The virtualities of the annihilating $\qb$ and the exchanged $g$ are of \order{Q^2}, hence the hard subprocess involves both quarks in the pion. This makes a longitudinal polarization of the photon possible, in contrast to the transverse polarization resulting from the $\qu\qb\to\gamma^*$ process with quarks of low virtuality.

As in the Bjorken limit we take $q^2=Q^2\to\infty$ with
\beq\label{xb}
x_B=\frac{q^+}{p^+}= \frac{Q^2}{2q\cdot p} = \frac{Q^2}{s}\ \ {\rm fixed}
\eeq
where we used \eq{xf} to set $q^- \simeq k^-$. The inclusive mass
\beq\label{my}
M_Y^2=(k+p-q)^2 \simeq (1-x_B)\left[s(1-x_F)+m_N^2 \right]-q_{\perp}^2
\eeq
being fixed in the BB limit \eq{bblimit} the momentum $k^-(1-x_F)$ of the ``stopped'' quark in the pion must be finite in the target rest frame. This quark remains coherent with the hard subprocess and its soft target interactions cannot be ignored, in analogy to the antiquark in the DIS process of \fig{dipole}. Thus we arrive at \fig{dybb}(b), which represents the amplitude for a specific state $Y$. (The inclusive DY cross section will be obtained below by squaring this amplitude and summing over $Y$.)

As the pion momentum in the target rest frame $k^-\to\infty$ and the relative transverse momentum $\kt$ of its quark constituents stays limited we may approximate the valence quark momenta as 
\beqa\label{ki}
k_1 &=& (0^+,zk^-,\kt) \nn\\
k_2 &=& (0^+,(1-z)k^-,-\kt)
\eeqa
The gluon $g(q_1)$ which transfers the $u$-quark momentum onto the $\bar d$ is highly virtual,
\beq\label{q1sq}
q_1^2 \simeq -zk^-\ell_1^+ \to -\infty
\eeq
which means that the pion Fock state is transversally compact and described by the pion distribution amplitude $\phi_\pi(z)$. Similarly
\beqa\label{qi}
q_2^2 &\simeq& -k^-\ell_1^+ \to -\infty \nn\\
q_1^- &\simeq& zq_2^- \simeq zk^- \to \infty
\eeqa
Hence the target vertices $\bar u(y_1)$ and $d(y=0)$ are separated by correspondingly short distances
\beqa\label{yi}
y_{1\perp} &=& \morder{1/Q} \to 0 \nn\\
{y}_{1}^+ &=& \morder{1/k^-} \to 0
\eeqa
On the other hand, the coherence length along the light-cone remains finite,
\beq\label{ioffe}
{y}_{1}^- = \morder{1/\ell_1^+}
\eeq

Noting that the hard subprocess in \fig{dybb}(b) is independent of $\ell_{1,2}^-$ and $\bs{\ell}_{1,2\perp}$ we see that the target blob is described by a GPD, in analogy to the DVCS case of \fig{DIS}(c) discussed above,
\beqa\label{amp9}
T(\pi^+ N\to\gamma^*_L Y) = \frac{-ieg^2\,C_F}{2\pi Q\sqrt{2N_c}}\int dx\, C(x_B,x)
\hspace{4cm} \\
\times \int dy_1^- \, e^{-iy_1^-l_1^+/2} \bra{Y(p')}\bar\psi_u(y_1)\gamma^+\gamma_5\,\psi_d(0) \ket{N(p)}_{y_1^+ = y_{1\perp}=0}\nn
\eeqa
where $x=\ell_1^+/p^+$. The matrix element corresponds to a ``transition'' ($N\to Y$) GPD and the hard subprocess gives at lowest order
\beq\label{Cdef}
C(x_B,x) \equiv  \int_0^1 dz\,\phi_\pi(z)  \left(\frac{e_u}{1-z}\inv{x_B+x +\ieps}+\frac{e_d}{z}\inv{x -\ieps}\right)
\eeq

\begin{figure}[h]
\centerline{\epsfig{file=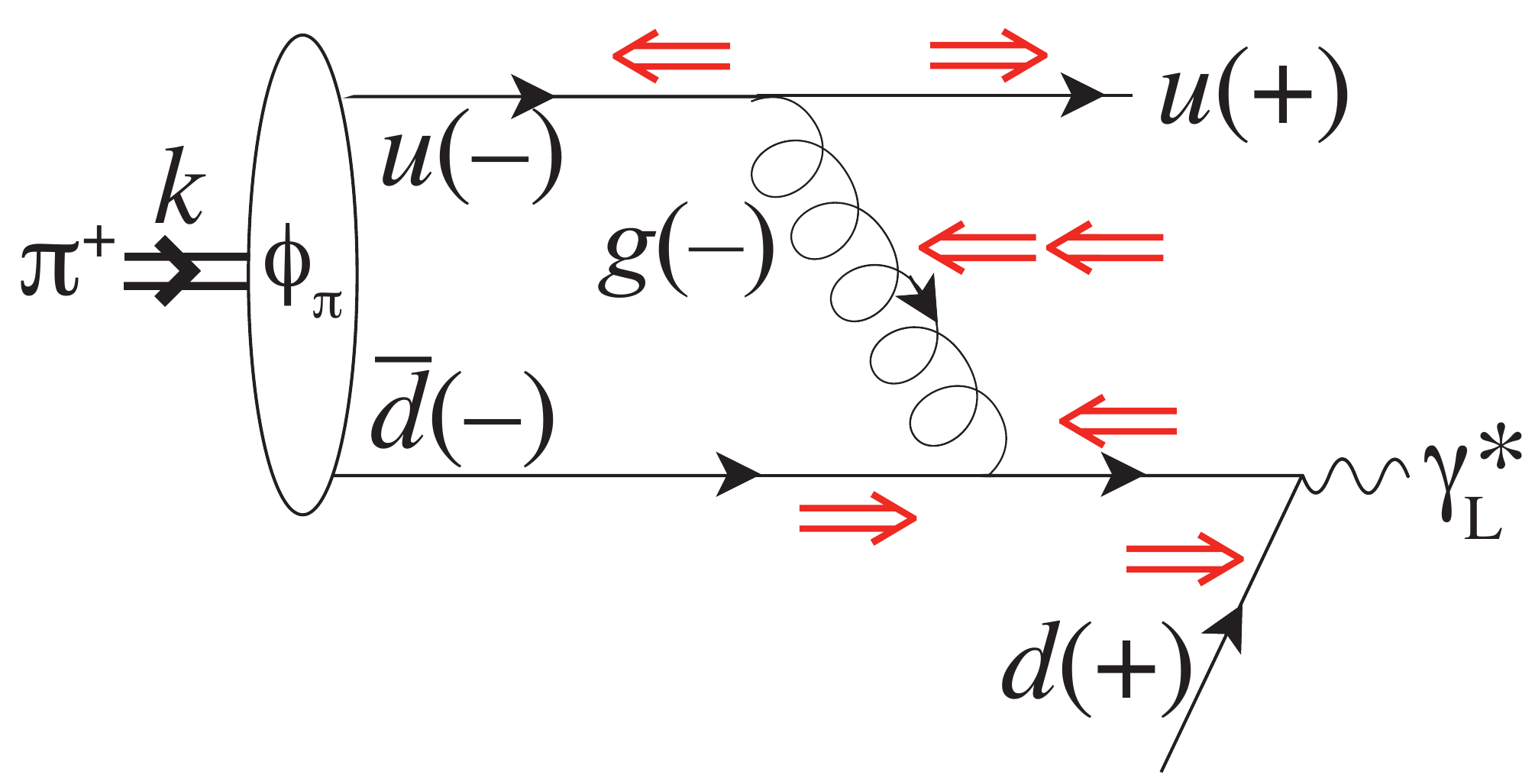,width=0.8\columnwidth}}
\caption{The double red arrows indicate the spin directions of the particles. All momenta are in the $\pm z$-direction as shown in the parentheses. The $u$-quark propagates into the GPD 
(not shown) while the $d$-quark propagates out of it.
\label{spin}}
\end{figure}
The virtual photon is dominantly longitudinal for any state $Y$ (we recall that $M_Y^2\ll s$ according to \eq{my}). In the case of $Y=N$ this is well-known from the time-reversed process $\gamma^* N \to \pi N$. The reason may more generally be understood as a consequence of the conservation of $J^z$ and the suppression of $L^z \sim q_\perp/Q$ due to the limited transverse momenta. In \fig{spin} the $u$-quark in the pion is taken to have helicity $+\halft$, hence $S_u^z=-\halft$ due to its large $k_1^-$ (as indicated in parenthesis, the $u$-quark is moving in the $-z$-direction). After emitting the gluon the $u$-quark has reversed its direction (more precisely, the hard process depends only on $\ell_1^+$ in \fig{dybb}(b)). Since its helicity is conserved it now has $S_u^z=+\halft$. The gluon takes up the difference, $S_g^z=-1$ since $L_g^z=\morder{1/Q}$. The $\bar d$ quark in the pion has $S_{\bar d}^z=+\halft$ since $S_\pi^z=0$. Helicity conservation then dictates that the target $d$-quark also has $S^z=\halft$ ($\bar d$ and $d$ have opposite helicity and move in opposite directions). After absorbing the gluon the $\bar d$ has $S_{\bar d}^z=\halft-1=-\halft$, and the photon thus gets $S_{\gamma^*}^z=-\halft+\halft=0$, {\it i.e.,} it is longitudinal. Since the $\bar d$ quark virtuality is of \order{Q^2} as it annihilates in the target its helicity and spin at that point are not simply related.

A corresponding analysis (as well as analytic calculation) shows \cite{Hoyer:2008fp} that the photon is transversely polarized in $p N \to \gamma^* Y$. In this case the Bj and BB limits give the same photon polarization, making it more difficult to distinguish the limits experimentally. The data on nucleon induced DY \cite{Zhu:2006gx} shows the photon to be transversely polarized in the full measured range of $x_F$.

\section{The inclusive $\pi^+ N \to \gamma^* Y$ cross section}

The DY amplitudes \eq{amp9} determine the inclusive cross section as
\beq\label{sig1}
\sigma(\pi^+ N\to\gamma^*_L Y) = \inv{2s}\sum_Y \int \frac{dq^- d^2\bs{q}_\perp}{(2\pi)^3 2q^-}\, |T(\pi^+ N\to\gamma^*_L Y)|^2 (2\pi)^4 \delta^4(k+p-q-p')
\eeq
The completeness sum
\beq\label{comp}
\sum_Y \ket{Y}\bra{Y} \equiv \sum_{n=0}^\infty\int\prod_{i=1}^n \frac{d^3\bs{p}_i}{(2\pi)^3 2E_i}\ket{\bs{p}_1,\ldots,\bs{p}_n} \bra{\bs{p}_1,\ldots,\bs{p}_n} =1
\eeq
requires an unlimited sum over all momenta $\bs{p}_i$ and thus also over the total momentum $\bs{p}'=\sum_i \bs{p}_i$, which is constrained by the momentum-conserving $\delta$-functions in \eq{sig1}. Before employing the closure relation we need to remove this explicit momentum constraint.

\begin{figure}[h]
\centerline{\epsfig{file=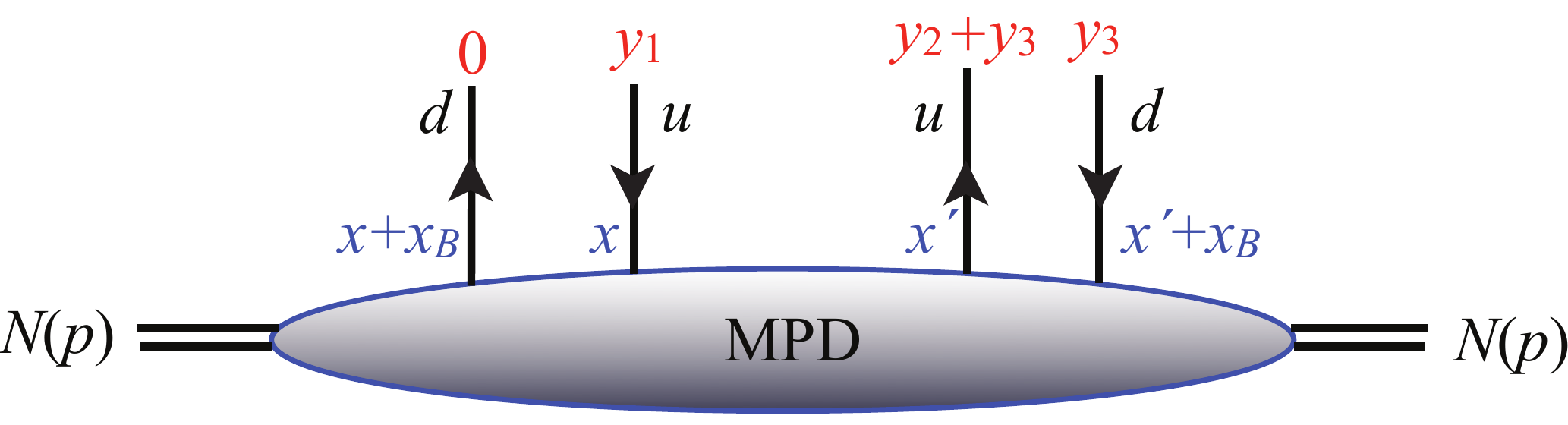,width=0.9\columnwidth}}
\caption{Pictorial representation of the forward multiparton distribution $f_{d\bar u/p}(x_B,x_M;x,x')$ given in \eq{dpdf}.
\label{MPD}}
\end{figure}

Integrating the inclusive cross section over the transverse momentum of the virtual photon eliminates the constraint on $\bs{p}_\perp'$. The longitudinal $\delta$-functions may be expressed as Fourier integrals, and the ${p'}^\pm$-dependent phase incorporated in the matrix element using translation invariance:
\beqa\label{shift}
\int dy_3^+\,dy_3^-\,\bra{N(p)}\bar\psi_d(0)\gamma^+\gamma_5\,\psi_u(y_2) \ket{Y(p')} e^{iy_3\cdot(k-q+p-p')} = \hspace{1cm}\nn\\
= \int dy_3^+\,dy_3^-\, \bra{N(p)}\bar\psi_d(y_3)\gamma^+\gamma_5\,\psi_u(y_2+y_3) \ket{Y(p')} e^{iy_3\cdot(k-q)}
\eeqa
The inclusive cross section is now seen to involve the multiparton distribution shown in \fig{MPD},
%
%
\beqa\label{dpdf}
f_{d\bar u/p}(x_B,x_M;x,x') = \hspace{8cm} (\theequation) \nn\\
=\frac{1}{4(4\pi)^3}
\int  dy_1^- dy_2^- dy_3^- dy_3^+ \exp\left\{\halft i\left[-y_1^-l_1^+ +y_2^-{l_1^+}'- y_3^-q^+ +y_3^+ x_Mp^-
\right]\right\} \nn\\ \nn\\
\times  \bra{N(p)}\bar\psi_d(y_3)\gamma^+\gamma_5\,\psi_u(y_2+y_3)\, \bar\psi_u(y_1)\gamma^+\gamma_5\,\psi_d(0) \ket{N(p)}_{y_{i\perp}=0;\ y_1^+=y_2^+=0}\nn
\eeqa\stepcounter{equation}
where $x'={\ell_1'}^+/p^+$ and
\beq\label{xm}
x_M = \frac{k^-(1-x_F)}{p^-}
\eeq
is the `$-$' momentum fraction of the pion (carried by the stopped quark in \fig{dybb}) which is transmitted into the inclusive system $Y$ and determines its mass,
\beq\label{xmass2}
M_Y^2 = m_N^2(1-x_B)(1+x_M)-\bs{q}_\perp^2 
\eeq
The constraint $M_Y \ge m_N$ implies
\beq\label{xMrange}
\frac{x_B+\bs{q}_\perp^2/m_N^2}{1-x_B} \leq x_M \le \infty
\eeq
In the BB limit $M_Y$ and hence $x_M$ are finite, and so is the conjugate variable, the LF time difference $y_3^+$ between the amplitudes $T$ and $T^\dag$.

The target matrix element is evaluated at both $y_3^+ > 0$ and $y_3^+ < 0$ in the MPD \eq{dpdf}, and is thus not LF time ordered. However, for $y_3^+ < 0$ the order of the $T$ and $T^\dag$ operators may be reversed by taking the hermitian conjugate of the matrix element. Hence the MPD may be expressed as the discontinuity of an LF time ordered matrix element, as expected from unitarity and indicated in \fig{DIS}(d). In this respect it differs from the multiparton distributions studied by Jaffe \cite{Jaffe:1983hp}, which give higher twist corrections to hard processes in the standard Bj limit. Those distributions are real, since all operators are evaluated at equal LF time.

Including the hard subprocess amplitudes the $\pi^+ N \to \gamma^* Y$ cross section in the BB limit is
\beqa\label{sig2}
\frac{d\sigma(\pi^+ N\to\gamma^*_L Y)}{dM_Y^2} =\hspace{8.5cm}\\ 
=\frac{2(eg^2C_F)^2}{Q^2 s^2(1-x_B)N_c}
\int dx\,dx'\, C(x_B,x) C^*(x_B,x')\, f_{d\bar u/p}(x_B,x_M;x,x')\nn
\eeqa
with $C(x_B,x)$ given in \eq{Cdef}.

\section{Reduction to incoherent jet production}

\begin{figure}[h]
\centerline{\epsfig{file=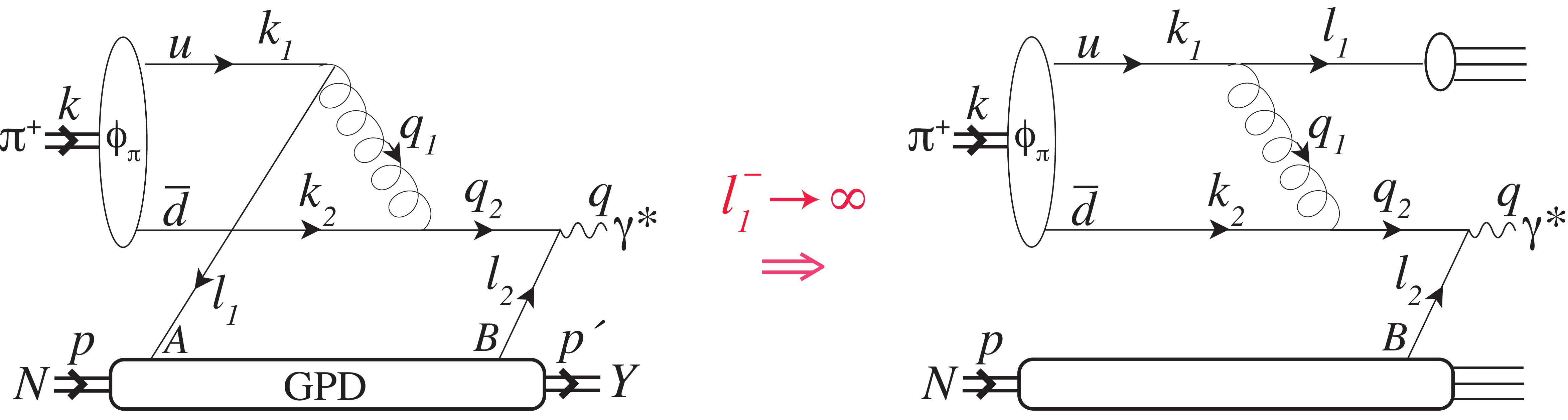,width=1.\columnwidth}}
\caption{In the secondary limit where the stopped $u$-quark momentum $\ell_1^-$ (and hence also $M_{Y}$) is large, the quark hadronizes independently of the target spectators (right). Squaring and summing the amplitudes turns the target matrix element into a $d$-quark PDF as in \eq{fdu2}.
\label{jet}}
\end{figure}

We have considered the BB limit \eq{bblimit} of the DY process in which $M_Y^2$ \eq{my} is fixed as $Q^2 \to \infty$. There was no restriction on the magnitude of $M_Y^2$. According to \eq{xmass2} the momentum $k^-(1-x_F)=x_M p^-$ of the `stopped' quark in \fig{dybb} grows with $M_Y^2$ and can become large enough for the quark to hadronize independently of the target spectators (\fig{jet}), as in the standard Bj limit. Thus for large $M_Y^2$ the MPD target matrix element \eq{dpdf} should reduce to a standard target PDF. This may indeed be verified as follows.

%
%

At large $x_M p^-$ the conjugate LF time vanishes, $y_3^+ \to 0$ in the MPD expression \eq{dpdf}. If we take the $u$-quark to be the stopped one as in \fig{dybb}, $\ell_1^-=x_M p^-$ is large and $\ell_1^+ \propto 1/x_M p^- \to 0$ in order that the $u$-quark remain close to on-shell. The $u$-quark propagation becomes light-cone dominated and the contraction of the fields $\psi_u(y_2+y_3)\bar\psi_u(y_1)$ may be approximated by the free propagator. We find \cite{Hoyer:2008fp}
\beq\label{fdu2}
f_{d\bar u/p}(x_B,x_M;x,x') \to \frac{1}{4\pi}\,\delta(x - x')\, \theta(x) f_{d/p}(x_B)
\hspace{1cm} (\ell_{1}^- \to \infty)
\eeq
The expression for the DY cross section in the BB limit at large $M_{Y}$ is then equivalent to the one obtained in \cite{Berger:1979du},
\beq\label{sig4}
\frac{d\sigma(\pi^+ N\to\gamma^*_L Y)}{dM_Y^2} = \frac{(ee_dg^2C_F)^2}
{Q^2 s^2(1-x_B)N_c}
\int \frac{dx}{2\pi x}\,\theta(x)\, \left(\int \frac{dz}{z}\phi_\pi(z)\right)^2\, f_{d/p}(x_B)
\eeq

\section{Summary}

The BB limit \eq{bblimit}, $Q^2\to\infty$ at fixed $(1-x_B)Q^2$, is not compatible with factorization at leading twist in DIS, $eN\to eX$. Coherence of the struck quark with the target remnant cannot be neglected at finite $M_X$ as is evident from the limiting case of $x_B=1$ ($eN\to eN$). Nevertheless, the success of Bloom-Gilman duality \cite{Melnitchouk:2005zr} indicates that both the Bj limit ($M_X\to\infty$) and the BB limit ($M_X=M_{N^*}$ fixed) are relevant for $eN\to eN^*$ (elastic and transition form factors).

The BB limit is appropriate for describing factorization in hard exclusive processes such as DVCS, $\gamma^*N \to \gamma(x_F)Y$, since a fixed mass $M_Y$ implies fixed $(1-x_F)Q^2$ (\cf\ \eq{my}). The coherence effects between the struck quark (which stops in the target after emitting the photon) and the target remnants are then described by the GPD.

We may use completeness in the system $Y$ to relate the cross section of the {\it inclusive} DVCS process $\sum_Y \sigma(\gamma^*N \to \gamma Y)$ to the discontinuity of a forward multiparton distribution (\fig{DIS}(d)). This applies to many other hard inclusive processes as well and offers novel opportunities to relate measurable cross sections to precisely defined target matrix elements.

We studied in detail the BB limit of the DY process $\pi N \to \gamma^*(x_F)Y$. This offers a possibility to estimate how high $x_F$ needs to be (or equivalently, how low the inclusive mass $M_Y$) for the BB limit to apply. In the Bj limit ($x_F$ fixed, $M_Y \to \infty$) the $\gamma^*$ is transversely polarized, whereas it is longitudinal in the BB limit ($x_F\to 1$, $M_Y$ fixed). For $Q^2 > 16\ \gev^2$ the transition was found  \cite{Anderson:1979xx} to start at $x_F \gsim 0.6$, with the $\gamma^*$ being dominantly longitudinal at $x_F=0.9$ ($M_Y \simeq 7$ GeV). In nucleon induced DY, $NN \to \gamma^*(x_F)+Y$, the $\gamma^*$ is predicted \cite{Hoyer:2008fp} to be transversely polarized in both the Bj and BB limits. The available data \cite{Zhu:2006gx} indeed shows no polarization change as a function of $x_F$.

The large Single Spin Asymmetries (SSA) observed at high $x_F$ in $p^\uparrow p \to \pi+Y$ \cite{Adams:1991rw} and $p p \to \Lambda^\uparrow+Y$ \cite{Bunce:1976yb} suggest another application of BB dynamics \cite{Hoyer:2006hu}. The SSA requires both helicity flip and a dynamic phase, both of which are suppressed in hard subprocesses. In the BB limit the helicity flip may occur in a {\it soft} interaction of a low-$x$ parton which is coherent with the hard process producing the hadron with high $x_F$ and $p_\perp$. This may explain the very large asymmetries observed as well as the puzzling fact that the asymmetry appears not to decrease with $p_\perp$, as expected in the standard leading twist framework.

\vspace{1cm}

{\bf Acknowledgements} I am grateful to the Organizers for their invitation to this meeting dedicated to the memory of Jan Kwieci\'nski, a close friend and collaborator of mine. The work described in this talk is based on my collaboration with Matti J\"arvinen and Samu Kurki. Travel support by the Magnus Ehrnrooth Foundation is gratefully acknowledged.

\end{document}